\documentclass[twocolumn,nofootinbib,prd]{revtex4}
\usepackage{amsmath}
\usepackage{graphicx}
\usepackage{amssymb}
\usepackage{mathrsfs}

\begin{document}

\title{Positivity and Hoop Properties for Chen-Nester-Tung Quasi-local Energy with Analytic Reference in Spherical Symmetry}

\author{Fei-hung Ho}\email{fei@ntnu.edu.tw}
\affiliation{
Department of Mathematics and Science, College of International Studies and Education for Overseas Chinese Students, National Taiwan Normal University, Taipei, Taiwan}
\affiliation{
Department of Physics, National Cheng-Kung University,
 Tainan, Taiwan
}

\author{Naqing Xie}\email{nqxie@fudan.edu.cn}
  \affiliation{School of Mathematical Sciences, Fudan University, Shanghai, China}

\date{\today}

\begin{abstract}
We provide a direct proof for the positivity of
Chen-Nester-Tung quasi-local energy with analytic reference in spherical symmetry . A hoop-type theorem for this energy is also established. Finally, the relation
between Chen-Nester-Tung and Brown-York quasi-local energies will be
discussed.
\end{abstract}

\pacs{04.20.Cv, 04.20.Fy}

\keywords{}

\maketitle

\section{Introduction}
In general relativity, at any point, one can always choose the coordinates so that the metric is the Minkowski metric with all first order derivatives vanishing. It is widely believed that the concept of the local energy density of a gravitational field is ill-defined and any attempt to localize the energy density must need a particular gauge. Nevertheless, a more modern idea, called `quasi-local',  arises and a good quasi-local energy has been long sought-after. These objects are defined associated with  a local density but rather with a closed 2-surface in a spacetime and one indeed tries to quantify the total energy inside the surface. There are many candidates for quasi-local energies among which each has certain desirable characteristics. A comprehensive survey is given by Szabados in \cite{Szabados}.

One particular formalism was developed by Chen-Nester-Tung's group \cite{CNT1}. This approach is applicable to a large class of geometric gravity theories and we will call this quantity the Chen-Nester-Tung quasi-local energy. Here let us briefly introduce their framework. Let $\text{\bf N}$ be a spacetime displacement vector field. The Hamiltonian 3-form has the general form
\begin{equation}
\mathcal{H}(\text{\bf N})=N^\mu \mathcal{H}_\mu+d\mathcal{B}(\text{\bf N})\end{equation}
where the 3-form $N^\mu \mathcal{H}_\mu$ is proportional to the field equation and thus vanishes `on shell'. Therefore , the Hamiltonian associated with a spatial region $\Omega$ is determined by the total differential boundary term,
\begin{equation}
E(\Sigma, \text{\bf N})=\int_\Omega \mathcal{H}(\text{\bf N}) =  \oint _{\Sigma=\partial \Omega} \mathcal{B}(\text{\bf N}).\end{equation}
Note that it depends only on the field value on the boundary $\Sigma=\partial \Omega$ and this value is indeed quasi-local.

One also introduces certain reference values which represent the ground state, i.e. the state having vanishing quasi-local quantities. For any quantity $\alpha$, we let $\bar \alpha$ be the reference value. The Chen-Nester-Tung boundary expression will contain terms  of the form $\Delta \alpha =\alpha -\bar \alpha$.

For Einstein's general relativity, the Chen-Nester-Tung preferred covariant-symplectic boundary term had been identified \cite{CNT1}:
\begin{equation}\label{BCNT}
\mathcal{B}(\text{\bf N})=\frac{1}{16\pi}(\Delta\Gamma^\alpha_{\ \beta}\wedge i_{\text{\bf N}}\eta_\alpha^{\ \beta}+\bar{D}_\beta N^\alpha \Delta \eta _\alpha^{\ \beta}) \end{equation}
where $\eta^{\alpha\beta}=\ast (\vartheta^\alpha \wedge \vartheta^\beta)$, $\Gamma^\alpha_{\ \beta}$ is the connection 1-form, $i_{\text{\bf N}}$ is the interior product with the vector field $\text{\bf N}$. This was also found at the same time by Katz, Bi\v{c}\'{a}k and Lynden-Bell by a Noether argument \cite{KBLB}.

We are particularly interested in the spherical isotropic coordinates in which
the physical spacetime metric has the form
\begin{equation}\label{mtc}
    ds^2=-N^2dt^2+\Phi^2(dr^2+r^2d\theta^2+r^2\sin^2\theta d\phi^2),
\end{equation}
where $N$ and $\Phi$ are assumed to be functions of the general time
and radial coordinates $t$ and $r$.

We make the choice of the reference by the analytic approach \cite{LCN}. Let the reference  be the flat Minkowski spacetime
analytically, i.e. just taking $N\equiv 1$ and $\Phi\equiv 1$. Thus, the quasi-local quantity has vanishing value for this reference.
We choose the $2$-surface $\Sigma$ to be a round sphere in a time slice,
i.e.
$
\Sigma=\{t=t_0, r=r_0\}.
$

Liu-Chen-Nester showed that \cite[Eqn. (A.17)]{LCN}
\begin{equation}\label{LCN}
    E_{\text{CNT}}(\Sigma,\text{\bf N})=\oint _{\Sigma=\partial \Omega}\mathcal{B}(\text{\bf N})=-r^2\Phi'(r)
\end{equation}
if we take
$
\text{\bf N}=e_\perp=\frac{1}{N}\frac{\partial}{\partial t}.
$
Here, $\prime$ denotes taking the derivative with respect to $r$.

\section{Positivity of $E_{\text{CNT}}(\Sigma, e_{\perp})$}
We show that, in spherical symmetry, Chen-Nester-Tung
quasi-local energy is nonnegative with analytic reference.

The induced $3$-metric of  time slice $\{t=t_0\}$ reads
\begin{equation}\label{3mtc}
    g^{(3)}=\Phi^2(t_0,r)(dr^2+r^2d\theta^2+r^2\sin^2\theta d\phi^2),
\end{equation}
which is conformally flat.

{\em Theorem:} If the scalar curvature $S(g^{(3)})\geq0$, then
$
E_{\text{CNT}}(\Sigma,e_{\perp})\geq0.
$

{\em Proof:} For technical reasons, we denote $\Phi^2=u^4$, i.e. $u=\sqrt{\Phi}$. It is well-known in conformal geometry that
\begin{equation}\label{Pos1}
    \Delta_{\mathbb{R}^3}u=\frac{1}{8}\Big(S(g^{\mathbb{R}^3})-u^4S(g^{(3)})\Big )u.
\end{equation}
Thus, the condition $S(g^{(3)})\geq 0$ implies that: for any $r_0>0$,
\begin{equation}\label{Pos2}
    \Delta_{\mathbb{R}^3}u=u''+\frac{2}{r}u'\leq0,
    \qquad\ \forall\, 0<r\leq r_0.
\end{equation}
Now we are trying to prove that for any $r_0$,
\begin{equation}\label{Pos3}
    -r_0^2\Phi'(r_0)\geq0.
\end{equation}
Since $\Phi'=2uu'$, it suffices to prove that $u '(r_0)\leq0$.
By Newton-Leibniz formula,
\begin{eqnarray}\label{Pos4}
    r_0u'(r_0) &=& \int^{r_0}_{0}(ru')'dr
    =\int^{r_0}_{0}(u'+ru'')dr \nonumber \\
                         &\leq& \int^{r_0}_{0}(-u')dr.
\end{eqnarray}
The last inequality comes from (\ref{Pos2}). One has
$r_0u'(r_0)\leq u(0)-u(r_0)
$
and finally we only need to show that
$u(0)-u(r_0)\leq0$
for any $r_0$.
Consider $u$ as a smooth function on $[0,r_0]$.
The strong maximum principle  in \cite[Theorem 3.5]{Gil-Tru} guarantees that the point where $u$ achieves its minimum must not be in the interior of $(0,r_0)$.
If it achieves its minimum at $0$, then by definition
$
u(0)-u(r_0)\leq0
$.
If it achieves its minimum at $r_0$, then
\begin{equation}\label{Pos6}
   u'(r_0)=\lim_{\delta\rightarrow0^+}\frac{u(r_0-\delta)
                      -u(r_0)}{-\delta} \leq 0.
\end{equation}
This is what we need.

{\em Remark:} The proof here is essentially a quasi-local version of Malec - \'{O} Murchadha inequality \cite[Eqn. (11)]{MOM}. Their result is working on the asymptotically flat manifolds while ours does not need the asymptotically flat condition near infinity. The proof here also reveals the spirit of Malec - \'{O} Murchadha.

\section{A hoop-type theorem}
If the energy is concentrated enough into a small region, gravitational collapse will happen. Thorne discussed this issue by focusing on the boundary geometry of the 2-surface. He proposed a hoop conjecture \cite{Thorne} which states:
\begin{quote} Horizons form when and only when a mass $M$ gets compacted into a region whose circumference in EVERY direction satisfies $C \lesssim 4 \pi M$. \end{quote}
He deliberately avoids defining either the 'circumference' or the mass. In spherical symmetry, it is reasonable to take $C=2\pi R$ where $R$ is the areal radius.

Recently, \'{O} Murchadha-Tung-Xie-Malec established a hoop-type theorem for the Brown-York quasi-local energy in spherical symmetry  \cite{OMTXM}. The spacetime is assumed to have a regular center and no past singularity. They show that if $C<2\pi E_{\text{BY}}$, then the surface must be trapped. Instead of being $4\pi$, the coefficient $2\pi$ here can be traced back to the fact that Thorn was considering something of the order of the Schwarzschild mass.

We also have the following hoop type theorem for Chen-Nester-Tung quasi-local energy:

{\em Theorem:}
If
$
C<2\pi E_{\text{CNT}}(\Sigma,e_{\perp}),
$
then $\Sigma$ is a trapped surface.

{\em Proof:}
By definition, the areal radius $R$ is
$
r\Phi(r)
$.
The condition $
C<2\pi E_{\text{CNT}}(\Sigma,e_{\perp})
$
implies that
$
    \Phi+r\Phi' < 0$.
The unit normal of $\Sigma$ in time slice is
\begin{equation}\label{UN}
    n^i=\left(\frac{1}{\Phi(r)},0,0\right).
\end{equation}
By a nice formula
\begin{equation}
k={n^i}_{;i}=\frac{1}{\sqrt{\text{det}g^{(3)}}}\left(n^i\sqrt{\text{det}g^{(3)}}\right)_{,i}
\end{equation}
in time slice, it is
$k=\frac{2(r\Phi'+\Phi)}{r\Phi^2}$ and is strictly negative.
Assuming spherical symmetry, the time-like expansion $p$ is a constant on the sphere. Let the null expansions be
$
\rho=(k+p)/\sqrt{8}
$ and
$
\mu=(k-p)/\sqrt{8}
$. If $p$ is positive, then $\mu=(k-p)/\sqrt{8} < 0$. If $p$ is negative, then $\rho=(k+p)/\sqrt{8} < 0$.
And if $p$ is zero, then both $\mu$ and $\rho$ are negative. If $\mu<0$, we know that there must be a past singularity. This is the possibility that we exclude. Therefore, we must have $\rho<0$ and $\mu>0$.
This turns out to be the original definition of a trapped surface by Penrose \cite{P}.

\section{The relation between Chen-Nester-Tung and Brown-York}
Recall that here we take the so-called analytic reference just by assigning $N\equiv1$ and $\Phi\equiv1$ so that the ambient spacetime becomes Minkowski.

However, there is also another way to take the reference spacetime to be Minkowski by isometric embedding. This is what Brown-York did \cite{Brown-York}.

Suppose that $\Sigma$ has positive Gauss curvature, then by Weyl theorem, $\Sigma$ can be isometrically embedded into $\mathbb{R}^3$, and $\mathbb{R}^3$ naturally lies in $\mathbb{R}^{3,1}$ with vanishing extrinsic curvature.

The Brown-York quasi-local energy is then defined as
\begin{equation}\label{BY}
    E_{\text{BY}}(\Sigma)=\frac{1}{8\pi}\int_{\Sigma}(k_0-k)d\Sigma,
\end{equation}
where $k$ and $k_0$ are the mean curvature of $\Sigma$ in the physical and reference space respectively. It should be noticed that in the analytic approach here, the induced $2$-metrics on $\Sigma$ are {\em not} isometric.
However, the Chen-Nester-Tung quasi-local energy does coincide with the Brown-York value.

The quasi-local quantity was calculated in Schwarzschild and it gave the standard values.
For instance, it recovers the Brown-York value \cite[Eqn. (A.20)]{LCN}
\begin{equation}\label{BYr}
    E_s(e_{\perp})=m\left(1+\frac{m}{2r}\right).
\end{equation}

Here we show that not only for Schwarzschild, $E_{\text{CNT}}(\Sigma,e_\perp)$ indeed recovers $E_{\text{BY}}(\Sigma)$ in spherical symmetry with $e_\perp=\frac{1}{N}\frac{\partial}{\partial t}$.

The proof is straightforward. We isometrically embed $\Sigma=\{t=t_0,r=r_0\}$ into $\mathbb{R}^3$. The mean curvature of the image in $\mathbb{R}^3$ is
\begin{equation}\label{kz}
    k_0=\frac{2}{r_0\Phi(r_0)}.
\end{equation}
Then
\begin{eqnarray}
  k_0-k &=& \frac{2\Phi(r_0)-(2r_0\Phi'(r_0)+2\Phi(r_0))}{r_0\Phi^2(r_0)} \nonumber \\
        &=& \frac{-2\Phi'(r_0)}{\Phi^2(r_0)}.
\end{eqnarray}
One has
\begin{eqnarray}
  E_{\text{BY}}(\Sigma)
   &=& \frac{1}{8\pi}\int_\Sigma(k_0-k)d\Sigma \nonumber \\
   &=& \frac{1}{8\pi}\frac{-2\Phi'(r_0)}{\Phi^2(r_0)}4\pi
       r_0^2\Phi^2(r_0) \nonumber \\
   &=& -r_0^2\Phi'(r_0) \nonumber \\
   &=& E_{\text{CNT}}(\Sigma,e_\perp).
\end{eqnarray}

This observation is not so obviously to find since the ways to choose the reference are quite different for Chen-Nester-Tung and Brown-York quasi-local energies. The former comes from the analytic approach while
the latter is via the isometric embedding.

\section{Conclusion and Discussion}
We have analyzed the positivity and hoop properties for both Chen-Nester-Tung and Brown-York quasi-local energies in spherical symmetry. These two energies are from different approaches of taking references but coincide with each other if we take the unit vector field
$
\text{\bf N}=e_\perp=\frac{1}{N}\frac{\partial}{\partial t}.
$ which is orthogonal to the constant `time' hypersurface.

Let $\{e_0,e_1,e_2,e_3\}$ be the orthonormal frame of the metric (\ref{3mtc}).
Working in the spherical symmetry, there is no contribution in the angular direction. For general $
\text{\bf N}=\text{N}^0e_0+\text{N}^1e_1
$, one yields
\begin{equation}\label{rst}
    E_{\text{CNT}}(\Sigma,\text{\bf N})=
    \text{N}^0E_{\text{CNT}}(\Sigma,e_0)=
    \text{N}^0E_{\text{BY}}(\Sigma).
\end{equation}
This may give an energy value either larger than or smaller than
the Brown-York.

If $E_{\text{CNT}}(\Sigma,\text{\bf N})>E_{\text{BY}}(\Sigma)$, then
$S(g^{(3)})\geq 0$ implies that $E_{\text{CNT}}(\Sigma,\text{\bf N})\geq 0$. The positivity follows directly from Shi-Tam's result \cite{ST}. If $E_{\text{CNT}}(\Sigma,\text{\bf N})<E_{\text{BY}}(\Sigma)$, then it satisfies the hoop property \cite{OMTXM}.

Recently, Chen-Liu-Nester-Sun proposed a new `best matched' reference approach to determine the preferred quasi-local boundary term \cite{CLNS}. It is required that the 4-metric be matching on the desired 2-surface and they extremize the associated quasi-local energy. In spherical symmetry, the `best matched' reference metric is
\begin{equation}d \bar s^2= -dT^2+dR^2+R^2 d\theta^2+R^2\sin^2\theta d\phi^2\end{equation}
with $T=N(t_0,r_0)t$, $R=\Phi(t_0,r_0)r$. For energy, the vector field should be the Killing field of the reference.  This is the same as $e_\perp$ for the surface $\Sigma=\{t=t_0, r=r_0\}$. Their `best matched' reference will give results equivalent to the Brown-York, and thus these results share the Shi-Tam positivity and hoop properties.

\begin{acknowledgments}
This work is partially supported by the National Natural Science Foundation of China (grants 11171328, 11121101)
and the Innovation Program of Shanghai Municipal Education Commission (grant 11ZZ01). The authors would thank James M. Nester for carefully reading the manuscript and many helpful comments. Part of this work was done while the first author was visiting School of Mathematical Sciences, Fudan University. He would thank the School for invitation, hospitality, and financial support.
\end{acknowledgments}


\begin{thebibliography}{}


\bibitem{Szabados}
L. Szabados, Living Reviews in Relativity {\bf 12}, 4 (2009).

\bibitem{CNT1}C.-C. Chang, J. M. Nester and C.-M. Chen, Phys. Rev.
Lett. {\bf 83}, 1897 (1999); J. M. Nester, Class. Quantum
Grav. {\bf 21}, S261 (2004); J. M. Nester, Mod. Phys. Lett. A {\bf 6}, 2655 (1991); C.-M. Chen, J. M. Nester and R.-S. Tung, Phys. Lett. A
{\bf 203}, 5 (1995); C.-M. Chen and J. M. Nester, Class. Quantum Grav. {\bf 16},
1279 (1999); C.-M. Chen and J. M. Nester, Grav. Cosmol. {\bf 6}, 257
(2000); C.-M. Chen, J. M. Nester and R.-S. Tung, Phys. Rev. D
{\bf 72}, 104020 (2005); J. M. Nester, Prog. Theor. Phys. Suppl. {\bf 172}, 30 (2008).

\bibitem{KBLB}D. Lynden-Bell, J. Katz, and Bi\v{c}\'{a}k, Mon. Not. R.
Astron. Soc. {\bf 272}, 150 (1995); J. Katz, J. Bi\v{c}\'{a}k and
D. Lynden-Bell, Phys. Rev. D {\bf 55}, 5957 (1997).
\bibitem{LCN}
J.-L. Liu, C.-M. Chen and J. M. Nester, Class. Quantum Grav. {\bf 28}, 195019 (2011).
\bibitem{Gil-Tru}
D. Gilbarg and N.S. Trudinger, {\it Elliptic Partial Differential Equations of Second Order} (Springer-Verlag Berlin Heidelberg, Germany, 2001).
\bibitem{MOM}
E. Malec and N. \'O Murchadha, Phys. Rev. D{\bf 50}, R6033 (1994).


\bibitem{Thorne}
K. S. Thorne in {\it Magic without Magic} ed. J. Klauder (Freeman, San Francisco), 231-258 (1972).

\bibitem{OMTXM}
N. \'{O} Murchadha, R.-S. Tung, N. Xie and E. Malec, Phys. Rev. Lett. {\bf 104}, 041101, (2010).
\bibitem{P}
R. Penrose, Phys. Rev. Lett. {\bf 14}, 57 (1965).

\bibitem{Brown-York}
J. D. Brown and J. W. York, Phys. Rev. D {\bf 47}, 1407 (1993).
\bibitem{ST}Y. Shi and L.-F. Tam, J. Diff. Geom. {\bf 62}, 79 (2002).

\bibitem{CLNS}  C.-M. Chen, J. M. Nester, J.-L. Liu and G. Sun, arXiv:1307.1510 (2013); G. Sun, C.-M. Chen, J. M. Nester and J.-L. Liu, arXiv:1307.1039 (2013).











\end{thebibliography}
\end{document}